\documentclass[%
 reprint,
superscriptaddress,
nofootinbib,
 amsmath,amssymb,
 aps,
prl,
]{revtex4-2}

\usepackage{graphicx}
\usepackage{dcolumn}
\usepackage{bm}

\usepackage{aas_macros}
\usepackage{xspace}
\usepackage{color}
\usepackage{hyperref}

\newcommand{\eg}{\mbox{e.\,g.,}\xspace}

\newcommand{\ie}{\mbox{i.\,e.,}\xspace}

\renewcommand{\vec}[1]{\bm{#1}}

\definecolor{darkgreen}{rgb}{0,0.5,0}
\definecolor{mypink1}{rgb}{0.858, 0.188, 0.478}
\definecolor{red}{rgb}{1,0,0}

\begin{document}

\preprint{APS/123-QED}

\title{Strong bound on canonical ultra-light axion dark matter from the Lyman-alpha forest}

\author{Keir K. Rogers}
 \email{keir.rogers@utoronto.ca}

\affiliation{%
Oskar Klein Centre for Cosmoparticle Physics, Department of Physics, Stockholm University,\\
AlbaNova University Center, Stockholm 10691, Sweden}%

\author{Hiranya V. Peiris}
 \email{h.peiris@ucl.ac.uk}

\affiliation{
Department of Physics \& Astronomy, University College London, Gower Street, London WC1E 6BT, UK}%

\affiliation{%
Oskar Klein Centre for Cosmoparticle Physics, Department of Physics, Stockholm University,\\
AlbaNova University Center, Stockholm 10691, Sweden}%

\date{\today}

\begin{abstract}
We present a new bound on the ultra-light axion (ULA) dark matter mass $m_\text{a}$, using the Lyman-alpha forest to look for suppressed cosmic structure growth: a 95\% lower limit $m_\text{a} > 2 \times 10^{-20}\,\text{eV}$. This strongly disfavors ($> 99.7\%$ credibility) the canonical ULA with $10^{-22}\,\text{eV} < m_\text{a} < 10^{-21}\,\text{eV}$, motivated by the string axiverse and solutions to possible tensions in the cold dark matter model. We strengthen previous equivalent bounds by about an order of magnitude. We demonstrate the robustness of our results using an optimized emulator of improved hydrodynamical simulations.\end{abstract}

\maketitle

\textbf{Introduction} -- The axion is a well-motivated dark matter particle candidate that can also explain the lack of observed CP violation in quantum chromodynamics \citep[the ``strong CP problem'';][]{PhysRevLett.38.1440, PhysRevLett.40.223, PhysRevLett.40.279, PRESKILL1983127, ABBOTT1983133, DINE1983137}. Ultra-light axions (ULAs) are axion-like particles with very small masses (\(m_\mathrm{a} \lesssim 10^{-10}\,\mathrm{eV}\)). These are generically produced in theories beyond the Standard Model, \eg string theories, which can predict the existence of many different axions \citep[the string ``axiverse''; \eg][]{2006JHEP...06..051S} that can comprise the dark matter \citep[\eg][]{2000PhRvL..85.1158H}. ULAs with masses \(\sim 10^{-22}\,\mathrm{eV}\) (also known as fuzzy dark matter) are of particular interest since this may be a preferred mass scale in the string axiverse \citep[\eg][]{2017PhRvD..95d3541H, 2019PhRvD..99f3517V}. Further, these axions are sufficiently light that wave-like behavior would manifest on astrophysical scales \citep[\(\sim\) kpc to Mpc;][]{2000PhRvL..85.1158H}. This could explain possible tensions in the standard cold dark matter (CDM) model between observations and simulations on galactic scales \citep[the so-called CDM ``small-scale crisis'';][]{2017ARA&A..55..343B}; although, \eg it is now clear that accurately simulating the co-evolution of dark matter and baryons is vital in this context \citep{2014Natur.506..171P}.

ULAs suppress the growth of cosmological structure below a certain scale (\(\sim\) Mpc). This scale is set by the so-called ``quantum pressure'' of ULAs \citep[\eg][]{PhysRev.85.166, PhysRevE.91.053304}. It is a function of axion mass such that heavier axions have smaller cut-off scales (at larger wavenumbers \(k\)). Current bounds from the early Universe exclude ULAs being more than half of the dark matter with masses \(m_\mathrm{a} \leq 10^{-23}\,\mathrm{eV}\) \citep{2015PhRvD..91j3512H, 2018MNRAS.476.3063H, 2015MNRAS.450..209B, 2015ApJ...803...34B, 2015PhRvD..91b3518S}.

In order to probe the canonical mass scale of \(10^{-22}\) eV, we must exploit the smallest scales currently accessible in the linear matter power spectrum. This is possible using the Lyman-alpha forest, neutral hydrogen absorption seen in high-redshift quasar spectra (\(2 \lesssim z \lesssim 6\)) \citep{1998ApJ...495...44C}. The absorption lines trace fluctuations in the intergalactic medium (IGM): the low-density (around mean cosmic density), largely primordial gas in-between galaxies. It follows that the flux power spectrum (correlations of the transmitted flux in the Lyman-alpha forest) is a powerful tracer of the linear matter power spectrum. By exploiting the highest resolution spectra available \citep{1994SPIE.2198..362V, bernstein2002volume, 2000SPIE.4008..534D}, we probe the matter power spectrum down to sub-Mpc scales \citep{2013PhRvD..88d3502V, 2017MNRAS.466.4332I, 2017JCAP...06..047Y, 2018ApJ...852...22W, 2019ApJ...872..101B} and hence power spectrum cut-offs from larger axion masses. The ULA smoothing ``Jeans'' length also mildly increases with redshift and so using higher-redshift Lyman-alpha forest measurements improves axion bounds.

In this work, we improve upon previous ULA bounds using the Lyman-alpha forest \citep{2017PhRvL.119c1302I, 2017MNRAS.471.4606A, 2017PhRvD..96l3514K} by exploiting a robust method for modeling the data, which we introduced in Refs.~\cite{2019JCAP...02..031R, 2019JCAP...02..050B}. This allows, for the first time, tests of the robustness of the bounds with respect to the fidelity of the theoretical modeling. ``Emulation'' of the flux power spectrum is necessary since ``brute-force'' sampling of the parameter space (as required by \eg Markov chain Monte Carlo methods; MCMC) is computationally infeasible due to the many expensive hydrodynamical simulations needed. An emulator is a computationally-cheaper but accurate model for the power spectrum, which can be called within MCMC and is built from a small set of ``training'' simulations \citep{rasmussen2003gaussian}. We optimize the construction of this training set by using Bayesian optimization \citep{10.1115/1.3653121}, a form of adaptive machine learning. The emulator model we use makes fewer assumptions and is more robust in its statistical modeling than existing linear interpolation techniques \citep[\eg][]{2017PhRvL.119c1302I, 2017PhRvD..96b3522I, 2020JCAP...04..038P}. More details on our methodology and cross-validation and convergence tests are presented in Ref.~\citep{2020RogersPRD}.

\textbf{Model} -- We model the effect of ULA dark matter on the IGM with suppressed initial conditions at \(z = 99\). This captures the small-scale power spectrum suppression that propagates to the flux power spectrum at \(z \sim 5\). The initial conditions are defined by a transfer function \(T(k) \equiv \left[\frac{P_\mathrm{ULA}(k)}{P_\mathrm{CDM}(k)}\right]^\frac{1}{2} = [1 + (\alpha (m_\mathrm{a}) k)^{\beta (m_\mathrm{a})}]^{\gamma (m_\mathrm{a})}\) \citep{2017JCAP...11..046M}. Here, \(P_\mathrm{ULA}(k)\) and \(P_\mathrm{CDM}(k)\) are respectively the linear matter power spectra for ULA dark matter and cold dark matter as a function of wavenumber \(k\). \(T(k)\) is characterized by three free functions \([\alpha (m_\mathrm{a}), \beta (m_\mathrm{a}), \gamma (m_\mathrm{a})]\), each a function of ULA mass \(m_\mathrm{a}\); \(\alpha(m_\mathrm{a})\) sets the scale of suppression, while \(\beta (m_\mathrm{a})\) and \(\gamma (m_\mathrm{a})\) set the shape of the power spectrum cut-off. We fit these functions using a polynomial model to transfer functions given by the modified Boltzmann code \texttt{axionCAMB}\footnote{\url{https://github.com/dgrin1/axionCAMB}.} \citep{Lewis:2002ah, 2017PhRvD..95l3511H}, which calculates cosmological evolution in the presence of a homogeneous ULA field. This parametric model accurately captures the key feature of a sharp small-scale cut-off in the power spectrum \citep[see][and also below]{2020RogersPRD}. Following Refs.~\cite{2019MNRAS.482.3227N, 2019PhRvD..99f3509L}, we model the effect of ULA quantum pressure only by modified initial conditions, as this is sufficient for the current sensitivity of data (see below).

For robust ULA bounds, we marginalize over uncertainties in the thermal state of the IGM. This accounts for suppression in the flux power spectrum arising from pressure smoothing, non-linear peculiar gas velocities and the thermal broadening of absorption lines \citep{2015MNRAS.450.1465G}. The vast majority of the IGM gas at about mean cosmic density (to which the forest is sensitive) at \(z \sim 5\) is well-described by a power-law temperature \(T(z)\) --- (over-)density \(\Delta\) relation \citep{1997MNRAS.292...27H}: \(T(z) = T_0(z) \Delta^{\widetilde{\gamma}(z) - 1}\). This has two free parameters: the temperature at mean density \(T_0 (z)\) and a slope \(\widetilde{\gamma} (z)\). We track the heat deposited in the IGM owing to cosmic reionization by the integrated energy injected per unit mass at the mean density \(u_0 (z)\) \citep{2016MNRAS.463.2335N}. This tracks the filtering scale in the IGM gas, which is the relevant pressure smoothing scale for an evolving thermal state in an expanding universe \citep{1998MNRAS.296...44G, 2015ApJ...812...30K}. The uniform ultra-violet (UV) photo-ionization rate is degenerate in the flux power spectrum with the mean amount of absorption in quasar spectra. We therefore account for uncertainty in the ionization state of the IGM by marginalizing over the effective optical depth \(\tau_\mathrm{eff} = - \ln \langle\mathcal{F}\rangle\), where \(\langle\mathcal{F}\rangle\) is the mean transmitted flux fraction. Our free parameter is a multiplicative factor \(\tau_0 (z = z_i)\) to the fiducial redshift dependence of \(\tau_\mathrm{eff}\) given by Ref.~\cite{2019ApJ...872..101B}.

In order to accurately bound the ULA power spectrum suppression scale, we marginalize over the slope \(n_\mathrm{s} \in [0.9, 0.995]\) and amplitude \(A_\mathrm{s} \in [1.2 \times 10^{-9}, 2.5 \times 10^{-9}]\) of the primordial power spectrum, with a pivot scale \(k_\mathrm{p} = 2\,\mathrm{Mpc}^{-1}\). Otherwise, we fix our cosmology to the baseline \textit{Planck} 2018 parameters \citep{refId0}: in particular, physical baryon energy density \(\Omega_\mathrm{b} h^2 = 0.022126\), physical dark matter energy density \(\Omega_\mathrm{c} h^2 = 0.12068\) and dimensionless Hubble parameter \(h = 0.6686\).

\textbf{Simulations} -- The 1D flux power spectrum measures correlations along the line-of-sight only (\ie integrated over transverse directions) in the transmitted flux \(\mathcal{F}\) normalized by the mean flux \(\langle\mathcal{F}\rangle\). In order to model this with sufficient accuracy, we run cosmological hydrodynamical simulations of the IGM using the publicly-available code \texttt{MP-Gadget}\footnote{\url{https://github.com/MP-Gadget/MP-Gadget}.} \citep{yu_feng_2018_1451799, 2001NewA....6...79S, 2005MNRAS.364.1105S}. We evolve \(512^3\) particles each of dark matter and gas in a \((10\,h^{-1}\,\mathrm{Mpc})^3\) box from \(z = 99\) to \(z = 4.2\). At each redshift bin of our data \(z = [4.2, 4.6, 5.0]\), we generate 32000 mock spectra (with pixel widths \(\Delta v = 1\,\mathrm{km}\,\mathrm{s}^{-1}\)) containing only the Lyman-alpha absorption line and measure the flux power spectrum using \texttt{fake\_spectra} \citep{2017ascl.soft10012B}.

Our simulations are optically thin, and heated and ionized by a spatially-uniform set of UV background (UVB) rates \citep{2012ApJ...746..125H}. In order to vary the output thermal IGM parameters \([T_0 (z = z_i), \widetilde{\gamma} (z = z_i), u_0 (z = z_i)]\) (see above), we vary the simulation input. This entails varying the amplitude \(H_\mathrm{A} \in [0.05, 3.5]\) and slope \(H_\mathrm{S} \in [-1.3, 0.7]\) in an overdensity-dependent rescaling of the default heating rates: \(\epsilon_i (z) = H_\mathrm{A} \epsilon_{0,i} (z) \Delta^{H_\mathrm{S}}\), for \(i \in [\mathrm{HI}, \mathrm{HeI}, \mathrm{HeII}]\). We also vary the mid-point redshift of hydrogen reionization \(z_\mathrm{rei} \in [6, 15]\) and the total heat injection during reionization \(T_\mathrm{rei} \in [1.5 \times 10^4, 4 \times 10^4]\,\mathrm{K}\) according to the model of Ref.~\cite{2017ApJ...837..106O}. Further discussion and tests of numerical simulation convergence and the effect of mis-modeling the mean flux (\ie using a rolling mean) is presented in Ref.~\cite{2020RogersPRD}.

\begin{figure}
\includegraphics[width=\columnwidth]{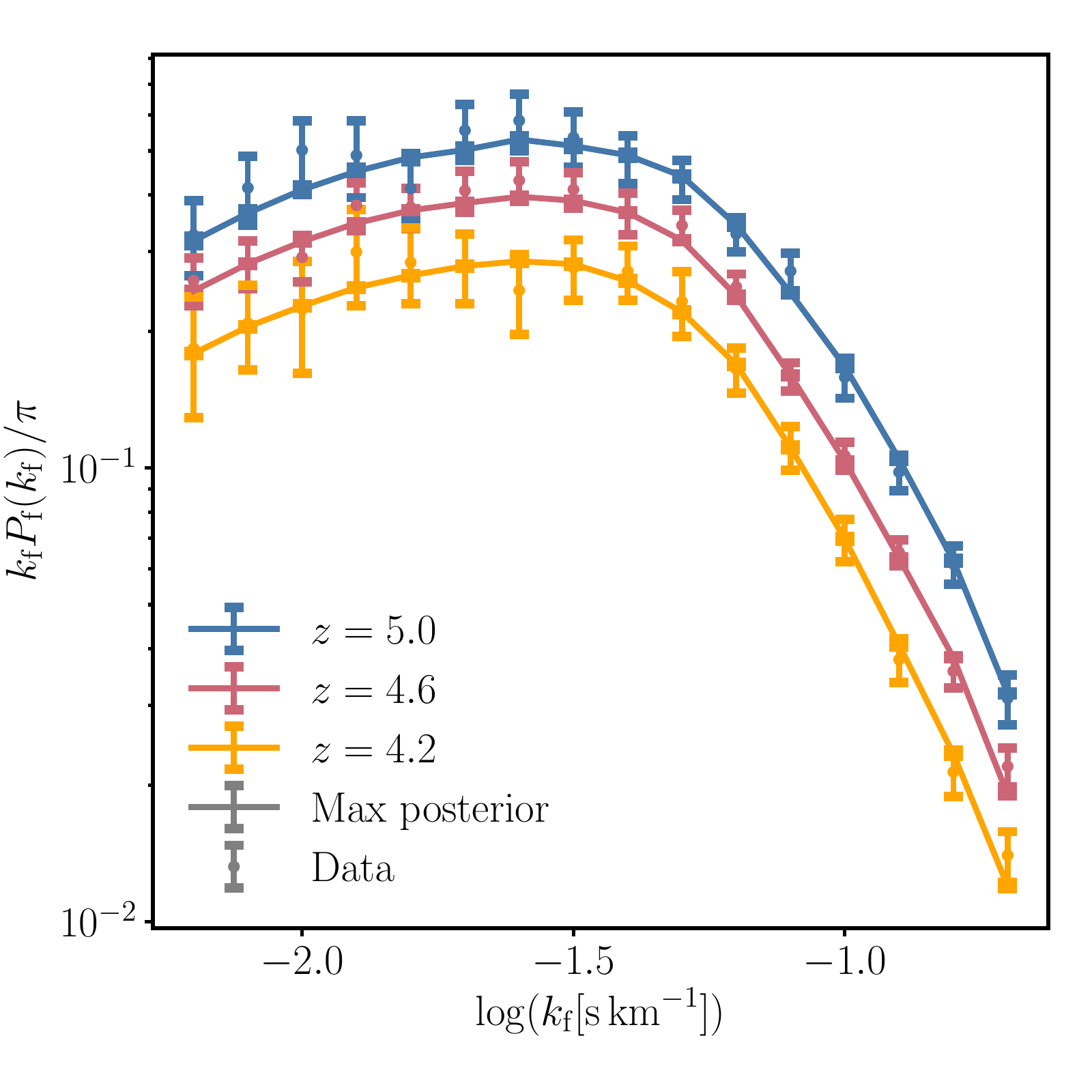}
\caption{\label{fig:data}A comparison of the 1D Lyman-alpha forest flux power spectrum \(P_\mathrm{f} (k_\mathrm{f})\) as measured by Ref.~\cite{2019ApJ...872..101B} and our maximum posterior model. Different colors show different redshifts \(z\) and \(k_\mathrm{f}\) is the line-of-sight velocity wavenumber.}
\end{figure}

\textbf{Data} -- We use the 1D flux power spectrum described in Ref.~\cite{2019ApJ...872..101B} and presented in Fig.~\ref{fig:data}. This includes smaller scales than previously accessed; we anticipate improved bounds on dark matter models as a consequence.

\textbf{Emulation and inference} -- In order to be able to sample the parameter space in a computationally-feasible manner, we ``emulate'' the flux power spectrum as a function of model parameters \(\vec{\theta} = [\alpha, \beta, \gamma, \tau_0 (z = z_i), T_0 (z = z_i), \widetilde{\gamma} (z = z_i), u_0 (z = z_i), n_\mathrm{s}, A_\mathrm{s}, \Omega_\mathrm{m}]\), separately at each redshift bin \(z_i = [4.2, 4.6, 5.0]\). We emulate as a function of \([\alpha, \beta, \gamma]\) (instead of \(m_\mathrm{a}\)) and fractional matter energy density \(\Omega_\mathrm{m}\) as part of our general emulator-inference framework for dark matter bounds \citep[see][]{2020RogersPRD}. We map from \(m_\mathrm{a}\) to \([\alpha, \beta, \gamma]\) using the parametric model defined above and fix \(\Omega_\mathrm{m} = 0.3209\) \citep{refId0}. We follow the Bayesian emulator optimization method we presented in Refs.~\cite{2019JCAP...02..031R, 2019JCAP...02..050B, 2020RogersPRD}, which uses a Gaussian process as the emulator model \citep{rasmussen2003gaussian}.

We adaptively optimize the construction of the emulator training set, ensuring convergence in parameter estimation with respect to the accuracy of the emulator model. In total, we build an emulator with 93 training simulations, each with ten samples evenly distributed in the \(\tau_0 (z = z_i)\) dimension at each redshift (since this parameter can be computationally-cheaply post-processed), \ie three emulators each with 930 training points. We present comprehensive tests of the method using cross-validation and convergence checks in Ref.~\cite{2020RogersPRD}; a summary is presented in the supplemental material.

We sample the posterior distribution for parameters \(\vec{\phi} = [\log(m_\mathrm{a} [\mathrm{eV}]), \tau_0 (z = z_i), T_0 (z = z_i), \widetilde{\gamma} (z = z_i), u_0 (z = z_i), n_\mathrm{s}, A_\mathrm{s}]\), for \(z_i = [4.2, 4.6, 5.0]\), using the MCMC ensemble sampler \texttt{emcee} \citep{2013PASP..125..306F}. We use a Gaussian likelihood function, with the data and their covariance as given above and the emulator covariance added in quadrature to propagate theoretical uncertainty. The theory flux power spectrum is predicted by the optimized emulator along with the modeling of the covariance at that position in parameter space.

In our prior distribution, we exclude the edges of the \(T_0 (z = z_i)\) --- \(u_0 (z = z_i)\) plane at each redshift not spanned by our training set, since these areas include unphysical IGMs, \ie high temperatures (high \(T_0\)) with little previous heating (low \(u_0\)) and \textit{vice versa}. Further, to prevent unphysical sudden changes in the IGM \citep[\eg][]{2017PhRvL.119c1302I, 2019ApJ...872..101B} in adjacent redshift bins (which would be inconsistent with previous observations \citep[\eg][]{2019ApJ...872...13W}), we prevent changes in \(T_0\) greater than 5000 K and changes in \(u_0\) greater than 10 \(\mathrm{eV}\,m_\mathrm{p}^{-1}\) (\(m_\mathrm{p}\) being the proton mass). We use \textit{Planck} 2018-motivated \citep{refId0} priors on \(n_\mathrm{s}\) and \(A_\mathrm{s}\) (translated to the pivot scale we use): Gaussian distributions respectively with means \(0.9635\) and \(1.8296 \times 10^{-9}\) and respectively standard deviations \(0.0057\) and \(0.030 \times 10^{-9}\). In order to disfavor very cold IGMs which are hard to motivate physically \citep[\eg][]{2017PhRvD..96b3522I} and following previous analyses \cite[\eg][]{2017PhRvL.119c1302I, 2017PhRvD..96l3514K, 2018PhRvD..98h3540M}, we use a conservative Gaussian prior on \(T_0 (z = z_i)\) with means set to our fiducial model ([8022, 7651, 8673] K at \(z = [5.0, 4.6, 4.2]\)) and standard deviations of 3000 K. As the effective optical depth is otherwise poorly constrained by our data and following Ref.~\cite{2018PhRvD..98h3540M}, we use a Gaussian prior on \(\tau_0 (z = z_i)\) with mean \(1\) and standard deviation \(0.05\).

Our prior is uniform in the logarithm of the ULA mass: \(\log(m_\mathrm{a} [\mathrm{eV}]) \in [-22, -19]\). This extends from the canonical ULA mass of \(10^{-22}\,\mathrm{eV}\) (which is already excluded in previous analyses) to the heaviest ULA mass that our data can probe (\(10^{-19}\,\mathrm{eV}\)); the power spectrum cut-off from heavier axions manifests on smaller scales than those accessible in our data.

\begin{table}
\begin{ruledtabular}
\begin{tabular}{lcc}
&\multicolumn{2}{c}{\(95 \%\) credible interval}\\
\hline
\(\log(m_\mathrm{a} [\mathrm{eV}])\)&\multicolumn{2}{c}{\(> -19.64\)}\\
\(n_\mathrm{s}\)&0.954&0.976\\
\(A_\mathrm{s}\)&\(1.77 \times 10^{-9}\)&\(1.88 \times 10^{-9}\)\\
\hline
\(\tau_0 (z = 4.2)\)&0.915&1.071\\
\(T_0 (z = 4.2)\) [K]&9334&12447\\
\(\widetilde{\gamma} (z = 4.2)\)&1.06&1.70\\
\(u_0 (z = 4.2)\) \([\mathrm{eV}\,m_\mathrm{p}^{-1}]\)&6.38&17.3\\
\hline
\(\tau_0 (z = 4.6)\)&0.951&1.062\\
\(T_0 (z = 4.6)\) [K]&9823&12838\\
\(\widetilde{\gamma} (z = 4.6)\)&1.18&1.60\\
\(u_0 (z = 4.6)\) \([\mathrm{eV}\,m_\mathrm{p}^{-1}]\)&7.24&17.2\\
\hline
\(\tau_0 (z = 5.0)\)&0.870&0.972\\
\(T_0 (z = 5.0)\) [K]&9195&11854\\
\(\widetilde{\gamma} (z = 5.0)\)&1.02&1.56\\
\(u_0 (z = 5.0)\) \([\mathrm{eV}\,m_\mathrm{p}^{-1}]\)&4.86&9.67
\end{tabular}
\end{ruledtabular}
\caption{\label{tab:bounds}1D marginalized \(95 \%\) credible intervals.}
\end{table}
\textbf{Results} -- Our main result can be summarized by a \(95 \%\) credible lower limit on the logarithm of the ULA dark matter mass (marginalized over the nuisance IGM and cosmological parameters described above): \(\log(m_\mathrm{a} [\mathrm{eV}]) > -19.64\), which equates to \(m_\mathrm{a} > 2 \times 10^{-20}\,\mathrm{eV}\). This is equivalent to a minimum allowed half-mode halo mass of \(7.2 \times 10^7\,\mathrm{M}_\odot\) \citep[\eg][]{2019arXiv190201055D}. Table \ref{tab:bounds} gives the full set of 1D marginalized \(95 \%\) credible intervals, while Fig.~\ref{fig:posterior} in the supplemental material shows a summary of the marginalized posterior distributions. Even when marginalizing over parameters which themselves induce small-scale suppression in the flux power spectrum (see above), the lightest axions that we consider (\(m_\mathrm{a} < \sim 10^{-20}\,\mathrm{eV}\)) are heavily disfavored relative to the cold dark matter limit. The IGM thermal state over which we marginalize is consistent with no statistically significant redshift evolution. This differs from some previous analyses which have suggested that the temperature at mean density increases from the highest to the next redshift bin \citep[\eg][]{2019ApJ...872...13W}; although others have also found no significant evolution \citep[\eg][]{2019ApJ...872..101B} and this is consistent with fiducial UVB heating rates \citep[\eg][]{2012ApJ...746..125H, 2019MNRAS.485...47P}. Further, we find no significant degeneracy between \(\log(m_\mathrm{a} [\mathrm{eV}])\) and \(T_0 (z = z_i)\) (see Fig.~\ref{fig:posterior}) owing to the wide range of scales and redshifts we exploit. The values of \(\widetilde{\gamma}\) and \(u_0\) are otherwise consistent (within \(95 \%\) limits) with previous observations \citep[\eg][]{2019ApJ...872...13W} and fiducial expectations \citep[\eg][]{1997MNRAS.292...27H}. The limits on the effective optical depth are consistent with our fiducial model \citep{2019ApJ...872..101B} except at \(z = 5.0\), where a lower optical depth is preferred (the marginalized mean is \(8 \%\) lower). The distributions on the cosmological parameters have not significantly updated from the prior, indicating as expected no constraining power on these parameters from our dataset.

Figure \ref{fig:data} compares the data we use (see above) with the maximum posterior flux power spectrum. The fit between data and model is good and the modeling uncertainties on the theory power spectrum are too small to be seen. This is because, due to the Bayesian emulator optimization (see above), the emulator uncertainty in the peak of the posterior (and its \(95 \%\) credible region) is smaller than the data uncertainty \citep[see][]{2020RogersPRD}.

\begin{figure}
\includegraphics[width=\columnwidth]{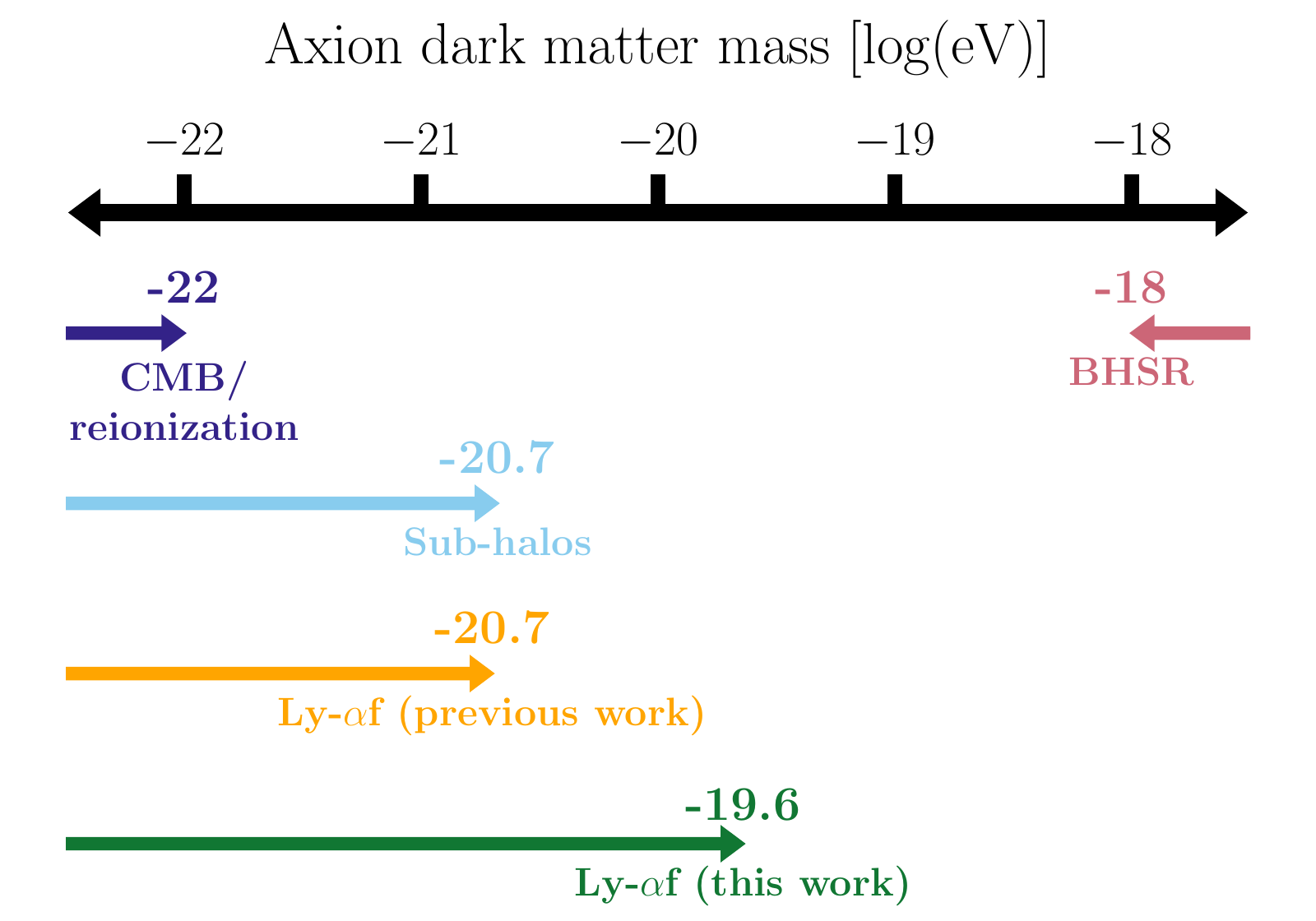}
\caption{\label{fig:comparison}Exclusion plot comparing our axion dark matter mass bound to other competitive bounds. ULA dark matter with masses \(10^{-33}\,\mathrm{eV} \leq m_\mathrm{a} \leq 10^{-24}\,\mathrm{eV}\) are excluded by \textit{Planck} \citep{2016A&A...594A...1P, 2018arXiv180706205P} cosmic microwave background (CMB) data \citep{2015PhRvD..91j3512H, 2018MNRAS.476.3063H}. A combination of the high-redshift UV luminosity function \citep{2015ApJ...803...34B} and the optical depth to reionization \citep{2015PhRvD..91b3518S} exclude at \(3 \sigma\) ULA dark matter for \(m_\mathrm{a} = 10^{-22}\,\mathrm{eV}\) (with some sensitivity to the reionization model) \citep{2015MNRAS.450..209B}. The non-detection of supermassive black hole super-radiance (BHSR) excludes \(10^{-18}\,\mathrm{eV} \lesssim m_\mathrm{a} \lesssim 10^{-16}\,\mathrm{eV}\) \citep{2011PhRvD..83d4026A, 2018PhRvD..98h3006S}. The sub-halo mass function excludes \(m_\mathrm{a} \lesssim 2.1 \times 10^{-21}\,\mathrm{eV}\) \citep{2020PhRvD.101l3026S}, while the equivalent previous bound (see main text) from the Lyman-alpha forest excludes \(m_\mathrm{a} < 2 \times 10^{-21}\,\mathrm{eV}\) \citep{2017PhRvL.119c1302I}. In this work, we exclude \(m_\mathrm{a} < 2 \times 10^{-20}\,\mathrm{eV}\) (at \(95 \%\) credibility). We consider here the case only where ULAs form all the dark matter; these bounds can be partially weakened if ULAs are a sub-dominant component.}
\end{figure}

\textbf{Discussion} -- Figure \ref{fig:comparison} compares our bound to some other competitive bounds (see caption for details). Our work closes a window of allowed ULA dark matter masses between the early Universe constraints at the lower end towards the black hole super-radiance bounds for higher masses. Our new lower limit on the ULA dark matter mass of \(2 \times 10^{-20}\,\mathrm{eV}\) improves over previous equivalent bounds \citep{2017PhRvL.119c1302I} by about an order of magnitude\footnote{Ref.~\cite{2017PhRvL.119c1302I} also report a slightly stronger ULA mass bound when setting a prior on a smoother thermal history; we compare to the more general result allowing jumps in the thermal history, which most closely matches our analysis.}. These bounds (including our own) can be weakened when considering the case where ULAs do not make up all the dark matter \citep{2017PhRvD..96l3514K}, but we defer analysis of these mixed dark matter models to future work.

We attribute the strengthening of the Lyman-alpha forest bound to key improvements in our analysis. First, we exploit data to much smaller scales (\(k_\mathrm{f}^\mathrm{max} = 0.2\,\mathrm{s}\,\mathrm{km}^{-1}\)) than in previous analyses \citep[\(k_\mathrm{f}^\mathrm{max} = 0.08\,\mathrm{s}\,\mathrm{km}^{-1}\);][]{2013PhRvD..88d3502V}. The ULA dark matter suppression scale (at which the 3D linear matter power spectrum drops by a half relative to CDM) \(k_\frac{1}{2} \propto m_\mathrm{a}^\frac{4}{9}\) \citep{2000PhRvL..85.1158H}. Although the mapping from the 3D power spectrum to the 1D flux power spectrum is non-trivial, we anticipate qualitatively that accessing more small-scale modes through the 1D power spectrum should improve the axion mass bound. We tested this: when we remove the smallest-scale (largest \(k_\mathrm{f}\)) bins from the data, the axion mass bound indeed weakens\footnote{For \(k_\mathrm{f}^\mathrm{max} = 0.126\,\mathrm{s}\,\mathrm{km}^{-1}\), \(\log(m_\mathrm{a} [\mathrm{eV}]) > -20.24\); for \(k_\mathrm{f}^\mathrm{max} = 0.08\,\mathrm{s}\,\mathrm{km}^{-1}\), \(\log(m_\mathrm{a} [\mathrm{eV}]) > -20.64\).}.

Second, we model simulated flux power spectra using a Bayesian-optimized Gaussian process emulator, which tests for convergence in parameter estimation with respect to the accuracy of the emulator model \citep{2020RogersPRD, 2019JCAP...02..031R, 2019JCAP...02..050B}. This contrasts with previous simulation interpolation methods \citep[\eg][]{2017PhRvL.119c1302I, 2017PhRvD..96b3522I, 2020JCAP...04..038P} which use Taylor expansion around a fiducial point and which we have shown in prior work can bias power spectrum estimation and weaken parameter constraints \citep{2019JCAP...02..050B}\footnote{Ref.~\cite{2018PhRvD..98h3540M} interpolate using ``ordinary kriging'', which has some similarity to our method, although they do not model parameter covariance nor test for convergence using Bayesian optimization.}.

Third, we marginalize over a physically-consistent IGM model (see above), which allows for a wide range of heating and ionization histories. In previous analyses, the temperature-density relation was varied freely as a function of redshift, along with a single redshift of reionization varied to trace the pressure smoothing in the IGM at all redshifts \citep{2017PhRvL.119c1302I, 2017PhRvD..96l3514K}. This means that IGMs were included with instantaneous temperatures \([T_0 (z), \widetilde{\gamma} (z)]\) inconsistent with the thermal history. Our model allows physically-motivated flexibility by additionally varying the total heat input during reionization \citep{2017ApJ...837..106O} and allowing for deviation from fiducial redshift dependencies in the integrated heating (by the \(u_0 (z = z_i)\) parameters). However, the exact impact on the axion mass bound from the balance between a more flexible IGM model and the physically-motivated priors that this allows is non-trivial. Finally, our prior is uniform in \(\log(m_\mathrm{a} [\mathrm{eV}])\), which we argue is less informative than in previous studies \citep[\eg][]{2017PhRvL.119c1302I}, where the prior is usually uniform in \(m_\mathrm{a}^{-1}\).

\textbf{Conclusions} -- We present a new lower limit on the mass of an ultra light axion dark matter particle at \(95 \%\) credibility: \(\log(m_\mathrm{a} [\mathrm{eV}]) > -19.64\) or \(m_\mathrm{a} > 2 \times 10^{-20}\,\mathrm{eV}\). This heavily disfavors (at \(> 99.7 \%\) credibility\footnote{The mass of \(10^{-21}\,\mathrm{eV}\) is disfavored in our analysis by \(\gg\) ``\(3 \sigma\)''; however, for robustness, we do not report bounds at \(> 99.7 \%\) credibility, since the tails of the distribution estimated by MCMC sampling can be unreliable.}) the canonical ultra-light axion with masses \(10^{-22}\,\mathrm{eV} < m_\mathrm{a} < 10^{-21}\,\mathrm{eV}\) as being the dark matter, motivated as a preferred mass scale in the string axiverse \citep{2017PhRvD..95d3541H, 2019PhRvD..99f3517V} and additionally, to solve the so-called cold dark matter ``small-scale crisis'' \citep{2017ARA&A..55..343B}. We have obtained this dark matter bound using the general emulator-inference framework we present in Ref.~\cite{2020RogersPRD}. In future work, we will exploit this framework to test other dark matter models, including mixed models where ULAs can be a subdominant component of the dark sector. There is further scope to extend the IGM model to include temperature and ionization fluctuations as a consequence of a spatially-inhomogeneous reionization \citep[\eg][]{2016MNRAS.460.1328D, 2019MNRAS.486.4075O, Suarez:2017xqg}, to which current data may be marginally sensitive \citep[\eg][]{2017PhRvD..95d3541H, 2019MNRAS.490.3177W}. Dark matter bounds can also benefit from upcoming Lyman-alpha forest observations, \eg from the Dark Energy Spectroscopic Instrument \citep{2016arXiv161100036D, 2016arXiv161100037D}, which can better determine the thermal and ionization state of the IGM. Our work additionally suggests that dark matter bounds could be further strengthened by accessing even smaller scales in the Lyman-alpha forest with higher-resolution spectroscopic observations.

\textbf{Acknowledgments} -- KKR thanks Jose O{\~n}orbe for sharing his code for the reionization model of Ref.~\cite{2017ApJ...837..106O} and for valuable discussions. The authors also thank Simeon Bird, George Efstathiou, Andreu Font-Ribera, Joe Hennawi, Chris Pedersen, Andrew Pontzen, Uros Seljak, Licia Verde, Risa Wechsler, Matteo Viel and the members of the \textit{OKC axions} journal club for valuable discussions. KKR thanks the organizers of the ``Understanding Cosmological Observations'' workshop in Benasque, Spain in 2019, where part of this work was performed. HVP acknowledges the hospitality of the Aspen Center for Physics, which is supported by National Science Foundation grant PHY-1607611. KKR was supported by the Science Research Council (VR) of Sweden. HVP was supported by the Science and Technology Facilities Council (STFC) Consolidated Grant number ST/R000476/1 and the research project grant ``Fundamental Physics from Cosmological Surveys'' funded by the Swedish Research Council (VR) under Dnr 2017-04212. This work was supported in part by the research environment grant ``Detecting Axion Dark Matter In The Sky And In The Lab (AxionDM)'' funded by the Swedish Research Council (VR) under Dnr 2019-02337. This work used computing facilities provided by the UCL Cosmoparticle Initiative; and we thank the HPC systems manager Edd Edmondson for his indefatigable support. This work used computing equipment funded by the Research Capital Investment Fund (RCIF) provided by UK Research and Innovation (UKRI), and partially funded by the UCL Cosmoparticle Initiative.

\subsection{\label{sec:supplement}Supplemental material}

\begin{figure*}
\includegraphics[width=0.247\textwidth]{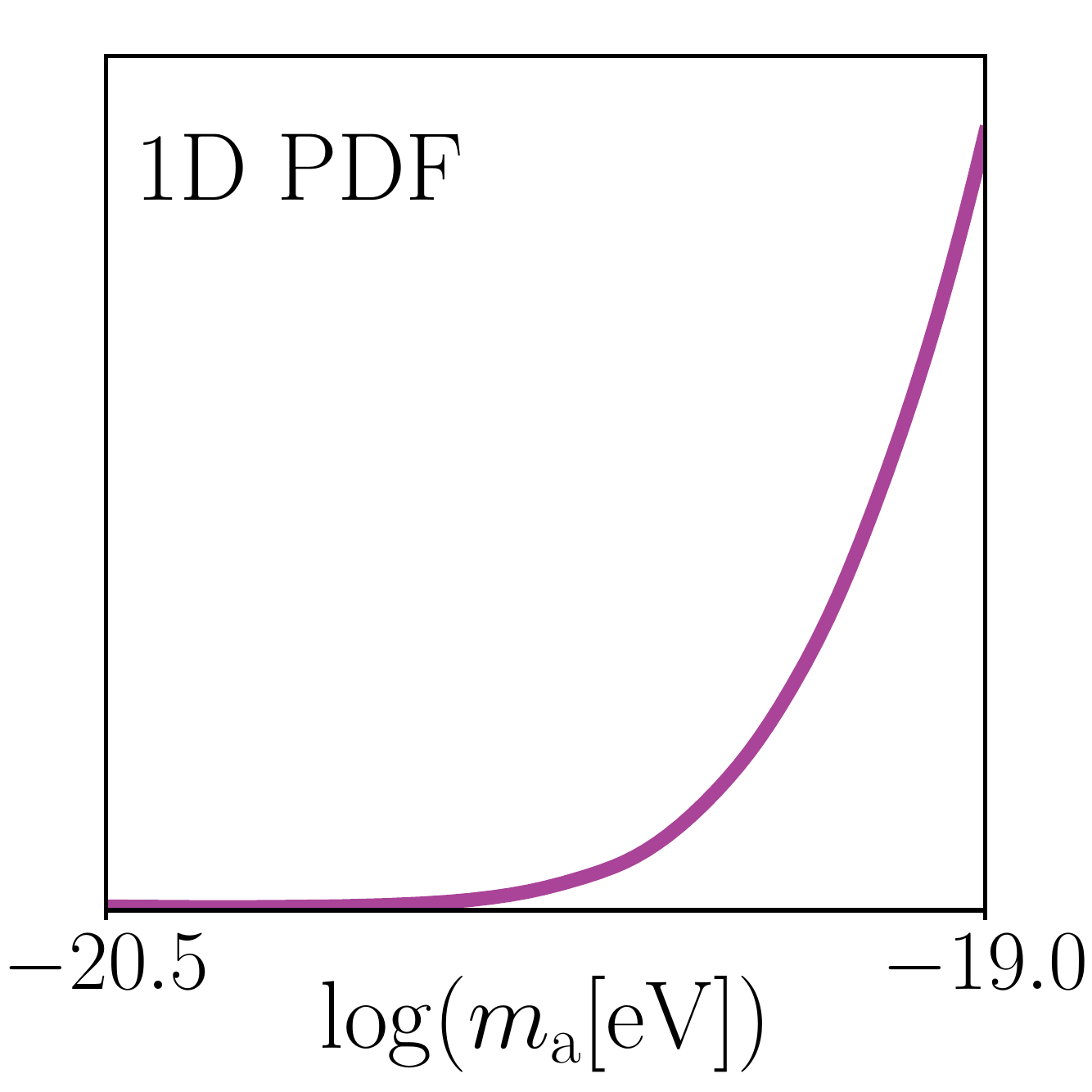} \includegraphics[width=0.74\textwidth]{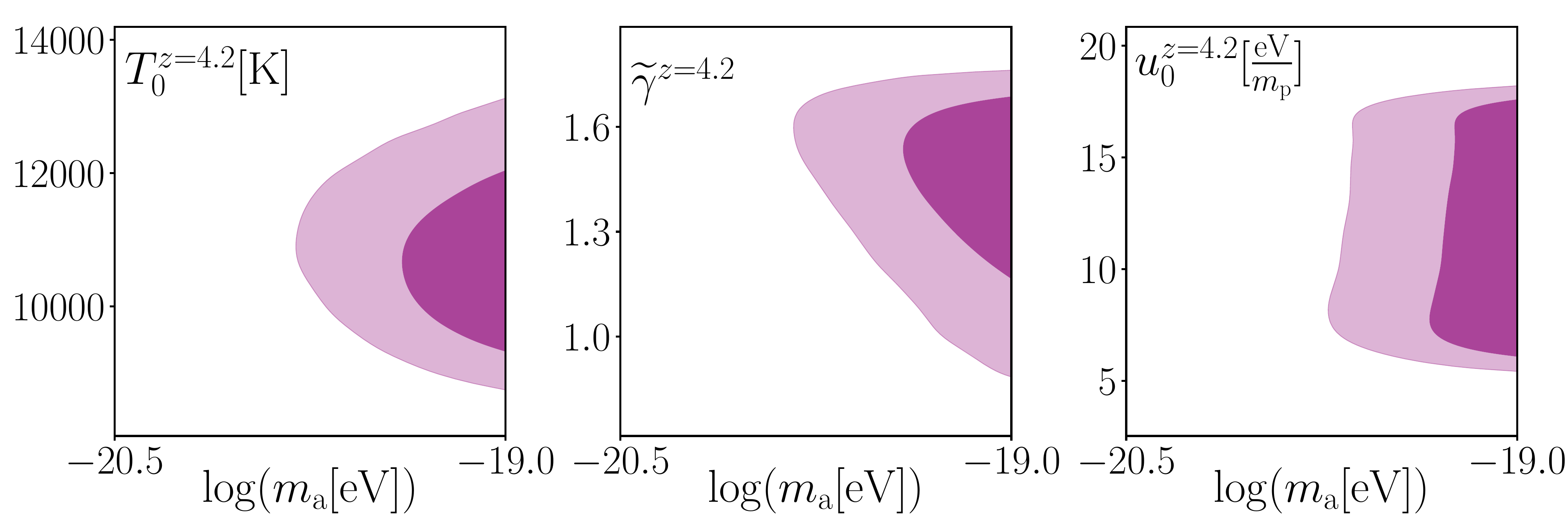}
\caption{\label{fig:posterior}A summary of our bound on the ultra-light axion dark matter mass \(m_\mathrm{a}\). The \textit{left} panel shows the 1D marginalized posterior. Then \textit{from left to right}, 2D marginalized posteriors with parameters describing the thermal state of the intergalactic medium at \(z = 4.2\): the temperature at mean cosmic density \(T_0\), the slope of the temperature-density relation \(\widetilde{\gamma}\) and the cumulative energy deposited per unit mass at the mean density \(u_0\). The darker and lighter shaded regions respectively indicate the \(68 \%\) and \(95 \%\) credible regions.}
\end{figure*}

\begin{figure}
\includegraphics[width=\columnwidth]{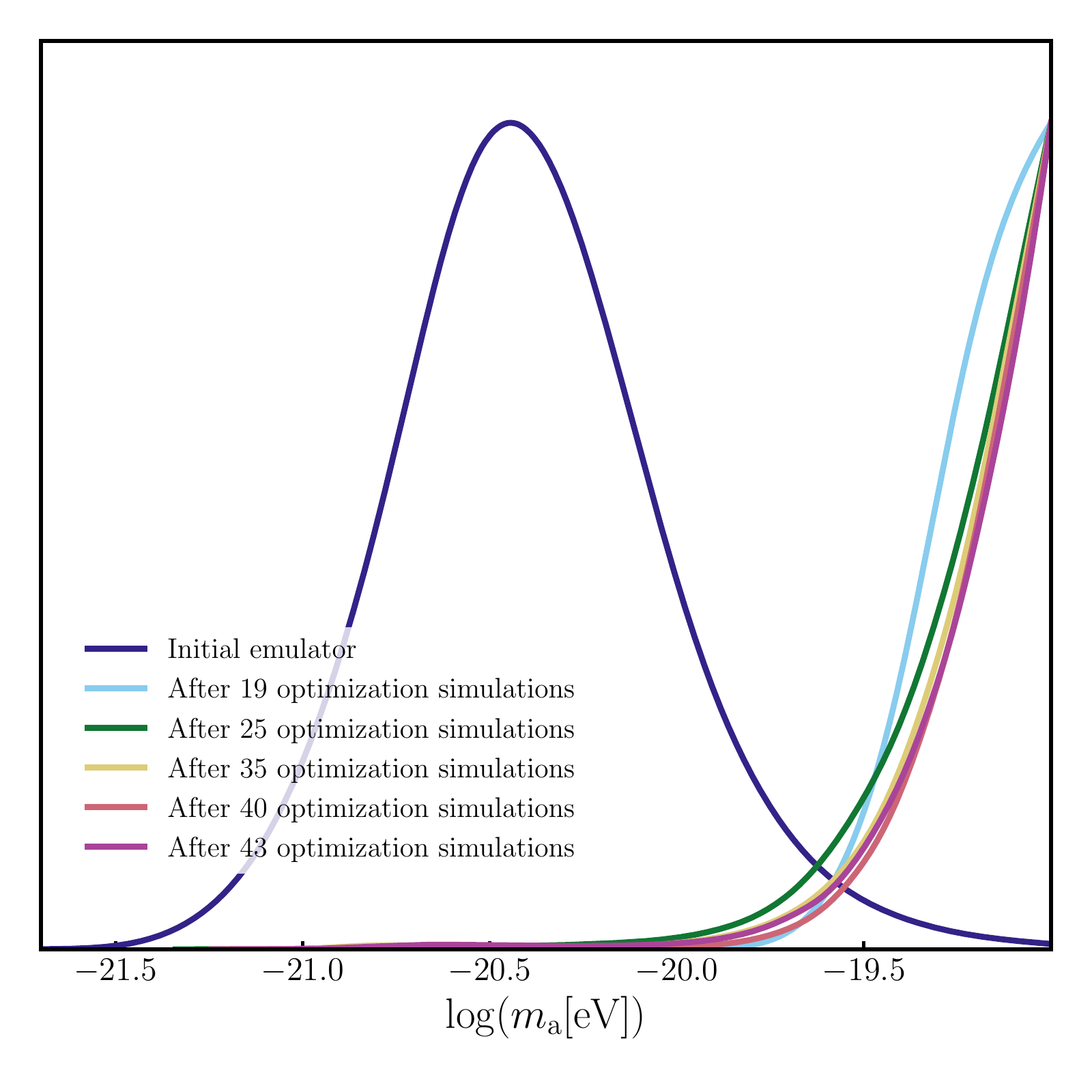}
\caption{\label{fig:posterior_logma}A summary of the convergence test for the Bayesian-optimized emulator: the 1D marginalized posterior for the logarithm of the ULA dark matter mass \(m_\mathrm{a} [\mathrm{eV}]\) at various iterations in the optimization of the emulator. Comprehensive tests of emulator convergence and cross-validation are presented in \citep{2020RogersPRD}.}
\end{figure}
We present here supplemental information on our results and methodology. Figure \ref{fig:posterior} shows a summary of the posterior distribution of the ULA dark matter mass and nuisance IGM parameters. Figure \ref{fig:posterior_logma} shows a summary of the emulator convergence tests presented in the companion article \citep{2020RogersPRD}. It shows estimates of the 1D marginalized posterior of \(\log(m_\mathrm{a} [\mathrm{eV}])\) at various iterations of the Bayesian-optimized emulator. After 19 optimization simulations have been added to the initial emulator of fifty, the bound on the axion mass varies little while a further 24 are added; and the convergence increases as more are added.

The full convergence test considers the marginalized posteriors of all parameters. Training data are added until successive estimates by MCMC of the posterior (marginalized mean and \(1 \sigma\) and \(2 \sigma\) constraints) converge with respect to each other. At each step, we maximize a modified GP-UCB acquisition function \citep{Cox97sdo:a, auer2002using, auer2002finite, dani2008stochastic}. This has two terms which balance \textit{exploitation} of approximate knowledge of the posterior (using the previous iteration of the emulator) to prioritize regions of high posterior probability, with \textit{exploration} of the full parameter space, prioritizing regions where modeling (emulator) uncertainty is high. The position in parameter space for each optimization simulation is the maximum acquisition point plus a random displacement. The size of the displacement is tuned to the \(95 \%\) credible region, in order to explore the peak of the posterior, which must be characterized most accurately. A second convergence criterion is also applied, checking that the exploration term of the acquisition function tends towards zero. This indicates that the acquisition is dominated by exploitation and so converged towards the posterior peak.

We propose training simulations in the space of inference parameters \(\vec{\phi}\), \ie we must then map from \(m_\mathrm{a}\) to \([\alpha, \beta, \gamma]\) and from \([T_0 (z = z_i), \widetilde{\gamma} (z = z_i), u_0 (z = z_i)]\) to \([H_\mathrm{A}, H_\mathrm{S}, z_\mathrm{rei}, T_\mathrm{rei}]\) to determine simulation input. For the first mapping, we use the model described in the main text and for the second, we interpolate a model using existing training data. We do not optimize in the \(\tau_0 (z = z_i)\) dimensions as these are sampled densely by post-processing. We used the ``batch'' version of the Bayesian optimization presented in Ref.~\cite{2019JCAP...02..031R}, which adds training data simultaneously in small batches from two to five; this made more efficient use of our computational resources.

\bibliography{axions}
\end{document}